\documentclass[conference]{IEEEtran}
\IEEEoverridecommandlockouts
\usepackage{amsmath,amsfonts}
\usepackage{algorithmic}
\usepackage{array}
\usepackage{textcomp}
\usepackage{stfloats}
\usepackage{url}
\usepackage{verbatim}
\usepackage{graphicx}
\usepackage{cite}
\usepackage{amssymb}
\usepackage{color,colortbl}
\usepackage{pgfplots}
\usetikzlibrary{calc}
\usepackage{multirow}
\usepackage{circuitikz}
\usepackage{subcaption}
\usepackage{mathtools}
\usepackage{algorithm}
\usepackage{algorithmic}

\usepackage{tikz}

\usepackage{environ}
\makeatletter
\newsavebox{\measure@tikzpicture}
\NewEnviron{scaletikzpicturetowidth}[1]{%
  \def\tikz@width{#1}%
  \begin{lrbox}{\measure@tikzpicture}%
  \BODY
  \end{lrbox}%
  \pgfmathparse{#1/\wd\measure@tikzpicture}%
  \BODY
}

\def\BibTeX{{\rm B\kern-.05em{\sc i\kern-.025em b}\kern-.08em
    T\kern-.1667em\lower.7ex\hbox{E}\kern-.125emX}}
\usepackage{pgfplots}
\pgfplotsset{compat=1.18}
\usepackage[normalem]{ulem}\usepackage{xspace}
\newcommand{\ed}[2][red]{\textcolor{#1}{#2}}

\usepackage[compatibility=false]{caption}
\begin{document}

\title{Deep Complex-valued Neural-Network Modeling and Optimization of Stacked Intelligent Surfaces}
\author{
    \IEEEauthorblockN{
    Abdullah Zayat,
    Omran Abbas, 
    Lo\"{i}c~Markley, 
    and Anas Chaaban }
    
    \IEEEauthorblockA{School of Engineering, University of British Columbia, Kelowna, Canada.\\ 
    Email: \{abdullah.zayat, omran.abbas, loic.markley, anas.chaaban\}@ubc.ca}
}

\maketitle

\begin{abstract}

We propose a {complex-valued neural-network} (CV-NN) framework to optimally configure {stacked intelligent surfaces} (SIS) in next-generation multi-antenna systems. Unlike conventional solutions that separately tune analog metasurface phases or rely strictly on SVD-based orthogonal decompositions, our method models each SIS element as a unit-modulus complex-velued neuron in an end-to-end differentiable pipeline. This approach avoids enforcing channel orthogonality and instead allows for richer wavefront designs that can target a wide range of system objectives, such as maximizing spectral efficiency and minimizing detection errors, all within a single optimization framework. Moreover, {by exploiting a fully differentiable neural-network formulation and GPU-based auto-differentiation, our approach can rapidly train SIS configurations for realistic, high-dimensional channels,} enabling near-online adaptation. Our framework also naturally accommodates hybrid analog-digital beamforming and recovers classical SVD solutions as a special case. Numerical evaluations under Rician channels demonstrate that CV-NN SIS optimization outperforms state-of-the-art schemes in throughput, error performance, and robustness to channel variation, opening the door to more flexible and powerful wave-domain control for future 6G networks.
\end{abstract}


\vspace{-0.2cm}

\section{Introduction}

With the continued evolution of wireless networks toward the sixth-generation (6G) era, researchers seek new paradigms that unify high-speed connectivity and pervasive intelligence \cite{6GIntelligentNetworkOfEverything}. The search for higher performance introduced techniques such as ultra massive MIMO (mMIMO), which inevitably escalates cost and energy usage \cite{Tataria,Chafii}, where even hybrid beamforming \cite{hybrid_beam_system} can become unwieldy.

To improve communication performance without incurring a large cost in terms of energy consumption, the idea of reconfigurable radio environments was proposed. This can be realized using reconfigurable intelligent surfaces (RIS) \cite{RIS_survey}, which passively reflect signals at tunable phases \cite{RISGen}. Since signals reflected through RIS suffer double path loss, it is favorable to place the RIS close to the transmitter or the receiver to improve performance. Moreover, it is desirable to have deeper wave-domain processing in such surfaces so as to enable functionalities needed in advanced 6G use cases.

To address this, stacked intelligent surfaces (SIS), commonly known as stacked intelligent metasurfaces \cite{activeSIM},\footnote{We use the term SIS here since the surfaces do not have to be metasurfaces in general.} recently emerged as a multi-layer reconfigurable surface approach, placed near transceiver antennas to impart sophisticated phase and/or amplitude control of impinging electromagnetic (EM) waves. By carefully tuning these layers, SIS can achieve complex wavefront manipulations such as beamforming, coding, or multiplexing directly in the electromagnetic wave domain, drastically reducing the RF-chain requirement and hardware complexity. Building on RIS concepts, SIS promises richer wave-domain capabilities for future 6G networks, including ultra-high data rates, pervasive AI integration, and advanced sensing \cite{SISISAC}.

Recent research on SIS span communications and sensing tasks. For instance, \cite{gradientdecentSIMComm} placed SIS at both transmitter and receiver to enhance holographic MIMO rates via wave-domain coding, while \cite{gradientdecentSIMSens} leveraged a multi-layer SIS for improved channel estimation in a multi-user MISO setting. In \cite{ISACSIM2}, SIS-assisted integrated communication and sensing (ISAC) was studied. Notably, many studies rely on optimizing the SIS using an SVD-based orthogonalization approach as in classical MIMO. Others adopt learning-based control (e.g., DRL \cite{DRL}), which can adapt SIS phases but often suffers from large exploration overhead and single-metric focus. Thus, a key challenge remains in efficiently configuring multi-layer SIS for dynamic, multi-objective 6G environments, where partial CSI, advanced sensing, and robust wave manipulation demand a more unified and scalable framework.

In this paper, we propose a novel perspective: treating SIS as a deep complex-valued neural network (CV-NN), where each meta-atom is a neuron with a unit-modulus weight. Wave propagation between SIS layers and between the SIS and antennas is embedded in fixed linear (complex) transformations, while the SIS phase shifts become trainable parameters analogous to neural weights. By doing so, standard deep-learning toolkits can perform backpropagation, allowing one to optimize arbitrary differentiable objective functions, such as sum-rate, detection error probability, or sensing accuracy. Moreover, this enables GPU-based auto-differentiation, which naturally scales to large metasurface arrays, sidestepping the overhead issues of DRL. SVD beamforming, partial iterative solutions, or specialized performance objectives can be realized as special cases of the proposed approach under carefully chosen objective functions. Overall, interpreting the SIS as a multi-layer CV-NN endows the system with the agility, multi-task potential, and real-time adaptability that 6G networks demand. We describe the modeling and optimization approach and demonstrate its superiority to SVD-based optimization via numerical simulations of symbol error rates.

The rest of this paper is organized as follows. Sec.~\ref{systemModel} outlines the SIS-assisted MIMO system. Sec.~\ref{sec.sisoptimizaiton} presents our CV-NN-based optimization method. Sec.~\ref{resultSection} provides numerical results benchmarked against conventional baselines, and Sec.~\ref{conclusion} concludes with future research directions.

\section{System Model}
\label{systemModel}

{We start by describing the considered MIMO system assisted by two SIS (one near the transmitter and one near the receiver). We then detail the channel model and the SIS-based transformations that shape the wavefront layer by layer.}

\subsection{System Description}
\begin{figure*}
\centering
 \includegraphics[width=1.1\linewidth]{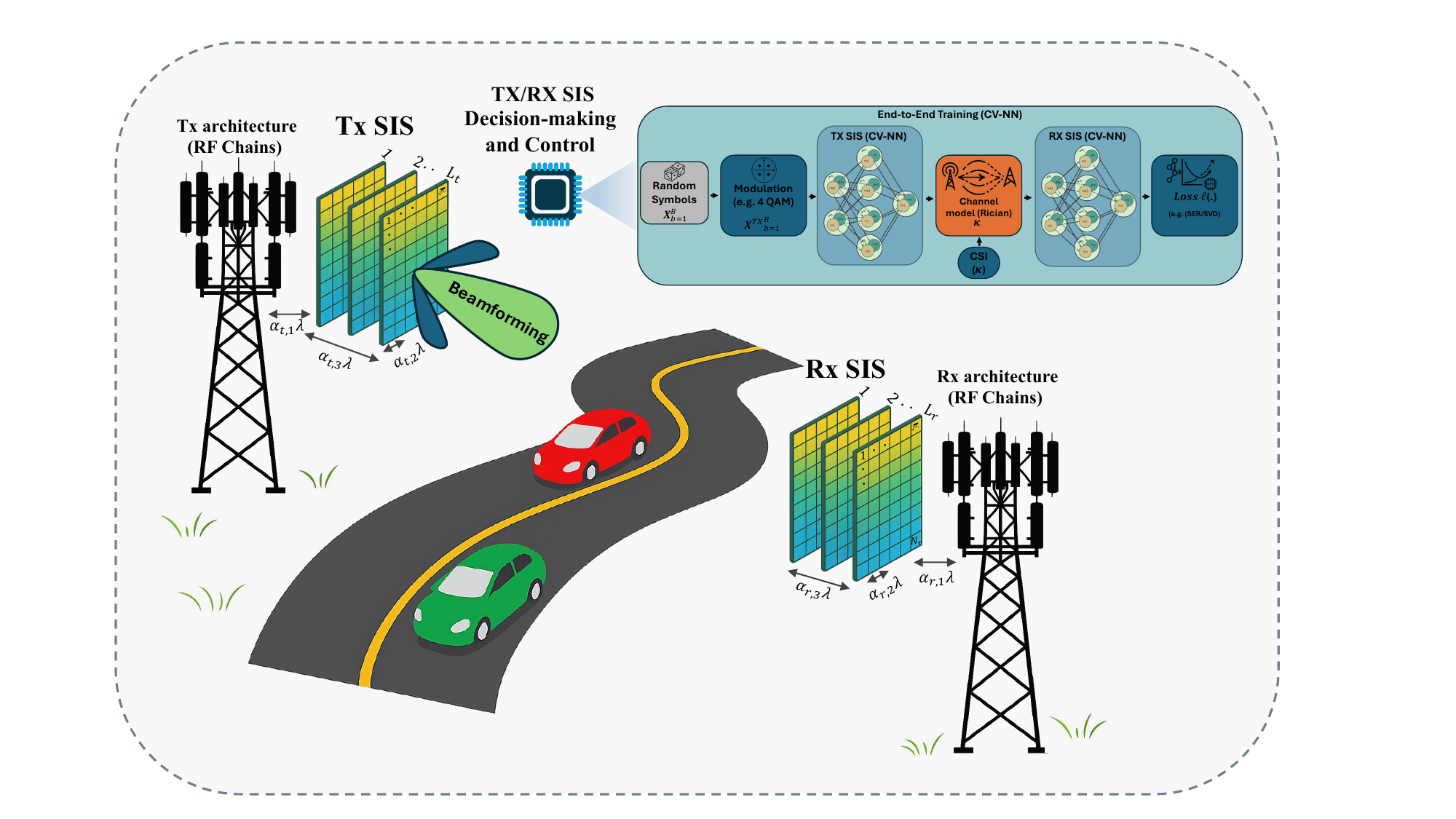}
\caption{{A MIMO system assisted by two stacked intelligent surfaces (SIS). Each SIS layer contains a 2D grid of programmable elements enabling transformations in the wave domain.}}
\label{joint_element}
\vspace{-0.5cm}
\end{figure*}

{
Fig.~\ref{joint_element} illustrates the considered MIMO system, wherein the transmitter (Tx) at $(x_{t},y_{t},z_{t})$ employs $M_{t}$ antennas and the receiver (Rx) at $(x_{r},y_{r},z_{r})$ employs $M_{r}$ antennas with $x_{r} > x_{t}$. We deploy a SIS close to each node to reduce hardware complexity at the transceiver. Specifically, the Tx-side SIS is placed at distance $\alpha_{t,1}\lambda$ from the Tx antenna array, has $L_{t}$ layers, and each layer comprises $N_{t}=N_{t,y}N_{t,z}$ programmable elements. The thickness of the SIS is $\alpha_{t,3}\lambda$, with inter-element spacing $\alpha_{t,2}\lambda$. 
Similarly, the Rx-side SIS has $L_{r}$ layers, each of $N_{r}$ elements, placed at distance $\alpha_{r,1}\lambda$ from the Rx. By adjusting each element's phase shift, the SIS can perform wave-domain beamforming or coding, ultimately redirecting or transforming the EM wave for improved link performance.}

{Let $l\in\{1,\dots,L_{t}\}$ index the Tx-SIS layers (with layer 0 denoting the Tx antenna plane), and let $i\in\{1,\dots,N_{t}\}$ index the elements. The 3D coordinates of element $i$ in layer $l$ follow $(x_{t}+\Delta_{x}^{(l)},\,y_{t}+\Delta_{y,i},\,z_{t}+\Delta_{z,i})$, where $\Delta_{x}^{(l)}$ accounts for the inter-layer spacing $\alpha_{t,3}\lambda/(L_{t}-1)$, and $\Delta_{y,i},\Delta_{z,i}$ capture horizontal offsets for the 2D grid. The Rx-SIS has a similar geometry, with $L_{r}$ layers and $N_{r}$ elements per layer. The SIS structure attains deep wave-domain processing power by applying phase shifts at each element across successive layers.}

\subsection{Channel and SIS Models}\label{channelModel}

We assume a Rician-fading channel $\mathbf{H} \in \mathbb{C}^{N_{r}\times N_{t}}$ between the {last} layer of the Tx-SIS and the {first} layer of the Rx-SIS {which remains constant for $T$ symbol (one block) and changes every block}. Under the far-field assumption, the channel coefficient $h_{i,j}$ from Tx-SIS element $i$ to Rx-SIS element $j$ can be written as
\begin{align}
    \label{eq.channelNEW}
    h_{i,j}= \frac{\lambda}{4\pi d}\left(\sqrt{\tfrac{\kappa}{\kappa +1}} e^{-2\pi j(\alpha_{t,2}\Delta_i + \alpha_{r,2}\Delta_j)} 
    + \sqrt{\tfrac{1}{\kappa+1}}l_{i,j}\right),
\end{align}
where $d$ is the distance between the SIS centers, $\kappa$ is the Rician factor, and $l_{i,j}\sim\mathcal{CN}(0,1)$ captures small-scale fading. The offsets $\Delta_i,\Delta_j$ reflect the physical 2D positions of the SIS elements. 

We define $\mathbf{P}\in\mathbb{C}^{N_{t}\times M_{t}}$ and $\mathbf{Q}\in\mathbb{C}^{M_{r}\times N_{r}}$ to represent the wave transformations from the Tx antennas to the output of the Tx-SIS and from the input of the Rx-SIS to the Rx antennas, respectively. These matrices are given by
\begin{subequations}
\label{eq:PQexp}
\begin{align}
    \mathbf{P} &= \mathbf{\Phi}^{L_{t}}\,\mathbf{W}^{L_{t}}\cdots \mathbf{\Phi}^{1}\,\mathbf{W}^{1},\\
    \mathbf{Q} &= \mathbf{E}^{L_{r}}\,\mathbf{\Theta}^{L_{r}}\cdots \mathbf{E}^{1}\,\mathbf{\Theta}^{1},
\end{align}
\end{subequations}
where $\mathbf{\Phi}^{l}$ and $\mathbf{\Theta}^{t}$ model the phase shifts imparted by the SIS Tx and Rx layers $l$ and $t$, respectively, and $\mathbf{W}^{1}$ and $\mathbf{E}^{1}$ model the propagation between the Tx and Rx SIS layers, respectively. Specifically, each Tx SIS layer $l$ is modeled by a diagonal matrix $\mathbf{\Phi}^{l}$ of dimension $N_{t}\times N_{t}$, containing the phase shifts $e^{j\phi_{l,i}}$ for $i\in\{1,\dots,N_{t}\}$. Likewise, the inter-layer propagation at the Tx SIS is modeled by a fixed matrix $\mathbf{W}^{l}$ (from layer $l-1$ to layer $l$, with layer $0$ denoting the Tx antennas) derived from the Rayleigh-Sommerfeld diffraction formula \cite{gradientdecentSIMComm}, where 
\begin{equation}
   \mathbf{W}^{l}_{i,i'} 
   = \frac{\alpha_{t,2}^{2}\lambda^{2}\cos\chi_{i,i'}}{r^{l}_{i,i'}}
   \left(\frac{1}{2\pi\,r^{l}_{i,i'}} - \frac{j}{\lambda}\right)e^{j2\pi r^{l}_{i,i'}/\lambda},
\end{equation}
where $r^{l}_{i,i'}$ is the distance between elements $i$ in layer $l-1$ and $i'$ in layer $l$, and $\chi_{i,i'}$ is the angle relative to the normal of the layer plane. At the Rx side, $\mathbf{E}^{l}$ is defined similarly, and $\mathbf{\Theta}^{l}$ is likewise a diagonal matrix comprising $e^{j\theta_{l,i}}$ phase shift terms.

\subsection{Signal Model and Objective}
Let $\mathbf{X}\in\mathbb{C}^{M_{t}\times T}$ denote the Tx baseband signals over $T$ symbols, and $\mathbf{Z}\in\mathbb{C}^{M_{r}\times T}$ be independent and identically distributed additive white Gaussian noise (AWGN) at the Rx with variance $\sigma^{2}$. After propagating through the Tx-SIS, the MIMO channel $\mathbf{H}$, and the Rx-SIS, the received signal is
\begin{equation}
    \label{eq:RxSignal}
    \mathbf{Y}
    =
    \mathbf{Q}\mathbf{H}\mathbf{P}\mathbf{X}
    +
    \mathbf{Z}.
\end{equation}
We primarily focus on a data-transmission scenario in which $\mathbf{X}$ contains modulated symbols (e.g., QAM or PSK), and the SIS is configured to minimize the symbol-error rate (SER). Nonetheless, the proposed approach could also incorporate other communication or sensing objectives. In all cases, the SIS phases $\{\phi_{l,i}\},\{\theta_{t,i}\}$ form the core variables to be optimized.

\section{SIS Optimization}
\label{sec.sisoptimizaiton}
{
By tuning each SIS element's phase shift, the multi-layer SIS can be jointly optimized alongside Tx/Rx processing to meet diverse performance objectives. Recently, simple SVD-based designs have been popular for capacity-like goals. Some iterative or learning-based methods also appeared in the literature but can become unwieldy for large SIS and time-varying channels. In what follows, we first briefly restate the conventional SVD approach, then we introduce our proposed CV-NN viewpoint. Finally, we show how the former paradigm may be treated as a special case of the latter, which forms a more general framework.}

\subsection{Conventional SIS Optimization}
\label{subsec:conventional_optim}
{
A well-known SIS design strategy attempts to realize or approximate the dominant singular modes of the end-to-end channel. Specifically, define the effective channel as}
\begin{equation}
\label{eq:Sigma_definition_2}
    \mathbf{\Sigma}
    \;\triangleq\;
    \mathbf{Q}\,\mathbf{H}\,\mathbf{P}.
\end{equation}
A classical approach is to {align} \(\mathbf{\Sigma}\) with the top singular modes of $\mathbf{H}$. Concretely, let
\begin{equation}
\label{eq:SVD_2}
   \widetilde{\mathbf{F}}\;\mathbf{\Pi}\;\widetilde{\mathbf{T}}^{H}
\end{equation}
represent the SVD (possibly truncated) of \(\mathbf{H}\) or \(\mathbf{\Sigma}\). An SIS-based solution then tries to realize
\begin{equation}
\label{eq:Sigma_approx_2}
\varphi \mathbf{\Sigma}\;\approx\;\mathbf{\Pi}_{1:M,1:M},
\end{equation}
denoting the top left $M\times M$ submatrix of $\Pi$, where \(\varphi\) is a constant scalar and $M=\min\{M_t,M_r\}$. Diagonalizing the channel in this way yields a nearly-diagonal channel matrix (orthogonal channels) with entries approximating the singular values \(\{\pi_{m}\}\) along the main diagonal of \(\mathbf{\Pi}\).
One can measure the alignment quality between $\mathbf{\Pi}$ and $\mathbf{\Sigma}$ via the normalized Frobenius gap
\begin{equation}
\label{eq:GammaC_2}
    \Gamma_{c}
    \;=\;
    \frac{\|\varphi\,\mathbf{\Sigma} \;-\;\mathbf{\Pi}_{1:M,1:M}\|_{F}^{2}}
         {\|\mathbf{\Pi}_{1:M,1:M}\,\mathbf{\Pi}_{1:M,1:M}^{H}\|_{F}^{2}},
\end{equation}
whose minimization forces \(\mathbf{\Sigma}\) to mimic the desired modes. Although effective for capacity-centric tasks, purely orthogonalization-based methods can be suboptimal under partial CSI or multiple objectives (e.g., joint detection and beamforming). Iterative schemes and reinforcement-learning approaches \cite{DRL} have been proposed, but they can become sample-inefficient or hard to scale with large surfaces. These limitations motivate a more flexible, end-to-end optimization framework.

\subsection{Proposed Complex-Valued Neural-Network Perspective}
\label{subsec:CNN_optim}

Rather than enforcing orthogonality or relying on a design that allows closed-form solutions, we propose to view SIS layers as a deep CV-NN \cite{CVNN}. In this formulation, each element in a layer plays the role of a ``neuron'' whose activation is a unit-modulus phase shift term. Concretely, a single Tx-SIS layer $l$ is modeled by a diagonal matrix $\mathbf{\Phi}^{l}$ acting on the wavefield \(\mathbf{X}_l\), with propagation between layers is captured by a fixed complex matrix \(\mathbf{W}^l\). Stacking \(L\) such layers yields an end-to-end transformation from input to output wavefields given by
\begin{equation}
\label{eq:CNNchain_2}
    \mathbf{X}_{l+1}
    \;=\; \bigl[\mathbf{X}_{l} \,\odot\, \mathbf{\Phi}^{l}\bigr]\,\mathbf{W}^{l},
\end{equation}
where \(\mathbf{X}_{l}\) denotes the wavefield at layer \(l\) and $\odot$ denotes the Hadamard product. We get similar for the transformation at the Rx-SIS with $\mathbf{\Theta}^{l}$ and $\mathbf{E}^{l}$ instead of $\mathbf{\Phi}^{l}$ and $\mathbf{W}^{l}$, respectively.
This structure is fully differentiable with respect to all SIS phases, enabling gradient-based updates which can be written for $\phi_{l,i}$ and $\phi_{l,i}$ as follows
\begin{subequations}
    \begin{alignat}{1}
        & \phi_{l,i}^{(t+1)} \;\leftarrow\; \phi_{l,i}^{(t)} \;-\; 
    \eta\,\frac{\partial\,\mathcal{L}}{\partial\,\phi_{l,i}}, \\
        & \theta_{l,i}^{(t+1)} \;\leftarrow\; \theta_{l,i}^{(t)} \;-\; 
    \eta\,\frac{\partial\,\mathcal{L}}{\partial\,\theta_{l,i}},
    \end{alignat}
\end{subequations}
where \(\eta\) is the learning rate and \(\mathcal{L}\) is any differentiable performance objective (e.g., symbol-error-rate). Modern deep-learning frameworks can automatically handle these derivatives, including the real and imaginary parts or using Wirtinger calculus \cite{Wirtinger}.

Moreover, any differentiable objective---from cross-entropy on detected symbols to radar contrast or multi-task criteria---can be plugged into this end-to-end pipeline. In essence, {the entire SIS becomes a trainable CV-NN}, enabling flexible, non-orthogonal beamforming and detection strategies. Such an approach can be readily implemented on standard GPU-friendly optimizers (e.g., Adam or RMSProp), opening the door to real-time adaptation under realistic 6G conditions.
 Note that classical SVD alignment effectively seeks to minimize the Frobenius gap \(\Gamma_c\) in \eqref{eq:GammaC_2}, thus enforcing near-orthogonal channels. Using an objective function that models \eqref{eq:GammaC_2} in our proposed approach can realize the classical SVD-style alignment as a special case. 

The advantage of our proposed CV-NN-based approach is that the SIS can learn any wave transformation, including non-orthogonal or multi-objective designs. Also, GPU-accelerated backpropagation handles large SIS arrays more efficiently than iterative or purely RL-driven methods. Hence, by interpreting each phase shift as a trainable parameter and each inter-layer propagation as a fixed complex kernel, the SIS optimization task becomes a generalized deep CV-NN problem, well-suited to 6G scenarios demanding real-time adaptation under partial CSI or multi-objective constraints.

\section{Results and Discussion}
\label{resultSection}

We now detail the end-to-end SIS training procedure and compare its performance with standard baselines. Specifically, we partition our presentation into three parts. First, we describe the experimental setup (modulation, channel, and training loop). Next, we benchmark the proposed approach against conventional closed-form and SVD-based solutions. Finally, we examine how SIS performance scales when varying the number of elements and layers.

\subsection{Experimental Setup}
We adopt a 4-QAM setup; therefore, {$\mathbf{X}$ in \eqref{eq:RxSignal} is independent identity disrupted symbols from 4-AOM.} The system operating frequency is $f = 3$~GHz, the transmitted power is set to $1$~dBm, the noise power is $\sigma^2 = -120$~dBm, and the Rician factor is \(\kappa = 15\). Accordingly, we use the following system parameters unless otherwise stated: The Tx has a linear antenna array along the z-axis with $M_{t} = 4$ at coordinates $(0,0,15)$. The Tx-SIS is placed at $4 \lambda$ from the antennas, with element spacing $\lambda/2$, thickness $6 \lambda$, number of layers $L_{t} = 3$, and number of elements $N_{t} = 484$. The Rx is located $50$ meters from the Tx and equipped with $M_{r} = 4$ antennas along the z-axis with Rx-SIS parameters similar to the Tx-SIS. Our end-to-end training loop generates random 4-QAM symbols in mini-batches and passes them through: 
\begin{enumerate}
\item the transmitter-side SIS layers, 
\item a stochastic Rician channel with \(\kappa=15\),
\item the receiver-side SIS layers.
\end{enumerate}
We then compute the symbol error rate (SER) via a cross-entropy loss as a surrogate objective function and backpropagate through all SIS layers (using PyTorch as a deep-learning toolkit). The SIS phases are updated using gradient descent under the unit-modulus constraint. This process is repeated until convergence. Algorithm~\ref{alg:E2Etraining} outlines these steps, from random symbol generation to the final SIS configuration. {When the training phase is finished, we fix the Tx-SIS and Rx-SIS parameters and test the SER. In the testing process, we find the SER from a new set of channel realizations by repeating steps 3 to 8 in Alg. \eqref{alg:E2Etraining} while freezing $\theta_{l,i}$ and $\phi_{l,i}$.}



\begin{algorithm}[!b]
\caption{\textbf{{End-to-End SIS Training }}}
\label{alg:E2Etraining}
\begin{algorithmic}[1]

\REQUIRE 
 \ed[white]{.} \\

\textbf{(a)}~Modulation (e.g.\ 4-QAM), 
\\
\textbf{(b)}~Tx/Rx antenna configs, 
\\
\textbf{(c)}~SIS params $(N_t,N_r,L_t,L_r)$, 
\\
\textbf{(d)}~Channel (Rician, factor $\kappa$),
\\
\textbf{(e)}~Loss \(\ell(\cdot)\) (e.g. cross-entropy (SER) or Frobenius gap (SVD).
\\

\ENSURE 
SIS phases \(\{\phi_{l,i}\}, \{\theta_{l,i}\}\) (unit-modulus).

\STATE \textbf{Initialize:} 
\(\phi_{l,i}, \theta_{l,i} \sim [0,2\pi)\). 
Set learning rate \(\eta\). 
Choose an SGD-based optimizer.

\FOR{epoch = 1 to \(\mathcal{E}\)}

  \STATE \textbf{(1)~Mini-batch:}
    Generate symbols \(\{\mathbf{X}_b\}_{b=1}^{B}\).

  \STATE \textbf{(2)~Forward pass:}
    \STATE \quad (a) 
      \(\mathbf{X}_{b}^{(\mathrm{TxSIS})}
       = \Bigl(\!\prod_{\ell=1}^{L_{t}} \mathbf{\Phi}^\ell \mathbf{W}^\ell\Bigr)\,\mathbf{X}_b.\)
    \STATE \quad (b) 
      Draw random \(\mathbf{H}_b\). 
      \(\;\mathbf{R}_b = \mathbf{H}_b \,\mathbf{X}_{b}^{(\mathrm{TxSIS})}.\)
    \STATE \quad (c)
      \(\mathbf{Y}_b
        = \Bigl(\!\prod_{\ell=1}^{L_{r}} \mathbf{E}^\ell \mathbf{\Theta}^\ell\Bigr)\,\mathbf{R}_b + \mathbf{Z}_b.\)

  \STATE \textbf{(3)~Compute loss:}\\
    \(\displaystyle
       \mathcal{L}
         = \sum_{b=1}^{B} \ell\bigl(\mathbf{Y}_b, \mathbf{X}_b\bigr),
    \)

  \STATE \textbf{(4)~Backprop:}
    \(\;
      \tfrac{\partial \mathcal{L}}{\partial \phi_{l,i}},\;
      \tfrac{\partial \mathcal{L}}{\partial \theta_{l,i}}.
    \)

  \STATE \textbf{(5)~Update phases:}
    \[\phi_{l,i} \leftarrow \bigl[\phi_{l,i} - \eta\,\tfrac{\partial\mathcal{L}}{\partial \phi_{l,i}}\bigr] \bmod 2\pi,\] \\
      \[\theta_{l,i} \leftarrow \bigl[\theta_{l,i} - \eta\,\tfrac{\partial\mathcal{L}}{\partial \theta_{l,i}}\bigr] \bmod 2\pi.\]

\ENDFOR

\STATE \textbf{Return:} \(\{\phi_{l,i}^*\},\,\{\theta_{l,i}^*\}\) minimizing \(\mathcal{L}.\)

\end{algorithmic}
\end{algorithm}

\subsection{Benchmarking vs.\ Closed-Form and SVD Approaches}
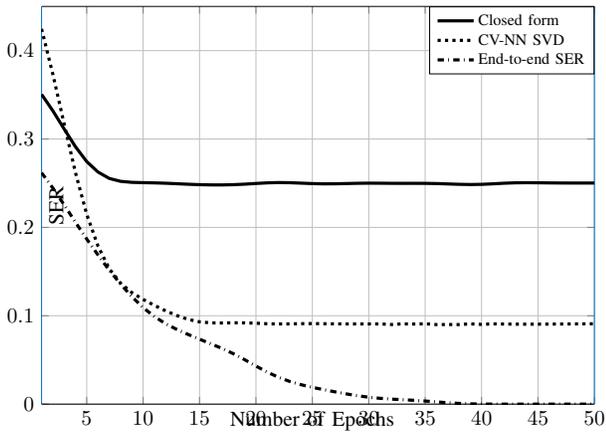
\begin{figure}
\centering
\tikzset{every picture/.style={scale=.8}, every node/.style={scale=1}}
%
%

\definecolor{mycolor1}{rgb}{0.00000,0.44700,0.74100}%
\definecolor{mycolor2}{rgb}{0.85000,0.32500,0.09800}%
\begin{tikzpicture}

\begin{axis}[%
grid=major,
width=4.521in,
height=3.255in,
scale only axis,
xmin=1,
xmax=50,
xlabel style={at={(axis cs:25,0)},anchor=north},
xlabel={Number of Epochs},
separate axis lines,
every outer y axis line/.append style={mycolor1},
ymin=0,
ymax= 0.45,
ylabel style={font=\color{black},at={(axis cs:1,0.225)},anchor=north},
ylabel={SER},
axis background/.style={fill=white},
xmajorgrids,
ymajorgrids,
legend style={{at =  (axis cs:50,0.45)}, nodes={scale=0.75, transform shape}, anchor=north east, legend cell align=left, align=left, draw=black}
]

\addplot [color=black, line width=1.5pt]
  table[row sep=crcr]{%
1	0.350519662640044\\
2	0.331841533808359\\
3	0.311342907755764\\
4	0.291284901528155\\
5	0.274460385287642\\
6	0.262632512972872\\
7	0.255382989711614\\
8	0.252106646993168\\
9	0.250947902503634\\
10	0.250604287119937\\
11	0.250350328430288\\
12	0.249944674849159\\
13	0.249377214210509\\
14	0.248797266952566\\
15	0.248327460297752\\
16	0.248057616104127\\
17	0.248032513867599\\
18	0.248309388686938\\
19	0.248892696486127\\
20	0.249671900159841\\
21	0.250377780822112\\
22	0.250711027476273\\
23	0.25054803995692\\
24	0.250052196919138\\
25	0.249536152163644\\
26	0.249254992282594\\
27	0.249305188953783\\
28	0.24958234046976\\
29	0.249882650008262\\
30	0.25001832786842\\
31	0.249965756164218\\
32	0.249849231491977\\
33	0.249807996059651\\
34	0.249832477172442\\
35	0.24980097915811\\
36	0.249577268674317\\
37	0.249161117962689\\
38	0.248714112962249\\
39	0.248486525455326\\
40	0.248686161237622\\
41	0.249271006945013\\
42	0.249959092474457\\
43	0.250428584132577\\
44	0.250563208727551\\
45	0.250466192395624\\
46	0.250313192276295\\
47	0.25020840278559\\
48	0.250166386598674\\
49	0.250165143179345\\
50	0.250177437140208\\
};
\addlegendentry{Closed form}

\addplot [color=black, dotted, line width=1.5pt]
  table[row sep=crcr]{%
1	0.424557034524204\\
2	0.373195780430209\\
3	0.317135569593066\\
4	0.262080724361858\\
5	0.214766305426501\\
6	0.17883378182572\\
7	0.153158311888355\\
8	0.13692601320166\\
9	0.126204114669753\\
10	0.118464980544332\\
11	0.111748407632654\\
12	0.105536328742883\\
13	0.100201970583332\\
14	0.0958289099791011\\
15	0.0931133116303671\\
16	0.0919711963827269\\
17	0.0918275462805642\\
18	0.0920694540808314\\
19	0.0919709214744666\\
20	0.0915401490619303\\
21	0.0909478400720686\\
22	0.0907461470857912\\
23	0.0907833273619711\\
24	0.0910597744977911\\
25	0.0911696641935905\\
26	0.0909778166190494\\
27	0.0908175652784049\\
28	0.0907508621980173\\
29	0.09074456181059486\\
30	0.0907419786720689\\
31	0.0907637046149047\\
32	0.0900738034285088\\
33	0.0908421847069821\\
34	0.0906204285869676\\
35	0.0909608735471263\\
36	0.09000751946748502\\
37	0.0900801240394363\\
38	0.0900376452095179\\
39	0.0908087811313666\\
40	0.0902155277664093\\
41	0.0908309412566183\\
42	0.0906216808027186\\
43	0.0904523327651545\\
44	0.0904364167490637\\
45	0.0904619417135198\\
46	0.0905453925412631\\
47	0.0906572455660871\\
48	0.0907589292842255\\
49	0.0908480026392636\\
50	0.0909549730495843\\
};
\addlegendentry{CV-NN SVD}

\addplot [color=black,dashdotted, line width=1.5pt]
  table[row sep=crcr]{%
1	0.261757678972925\\
2	0.244490752414655\\
3	0.225504811217293\\
4	0.205837646062698\\
5	0.186812164528049\\
6	0.168770096026056\\
7	0.151480536734914\\
8	0.135856668759801\\
9	0.12162892417768\\
10	0.109252345471183\\
11	0.0990592293125988\\
12	0.090955074599149\\
13	0.0844514444945246\\
14	0.078773163877126\\
15	0.0734875637386549\\
16	0.0682650320188574\\
17	0.0630176361221946\\
18	0.0570638146779182\\
19	0.0502765402822328\\
20	0.0431056875998911\\
21	0.0362350540363048\\
22	0.030361460797299\\
23	0.0256382163865102\\
24	0.0219861843498493\\
25	0.0189865263204955\\
26	0.0163880245935899\\
27	0.0139009190964678\\
28	0.0116002757335934\\
29	0.00953112893303183\\
30	0.00781892146251027\\
31	0.00656247256030616\\
32	0.00567466271195751\\
33	0.00499821706520529\\
34	0.00431847289169378\\
35	0.00349675420070317\\
36	0.00252222643769675\\
37	0.00156992511238706\\
38	0.000816668669281405\\
39	0.000345609766184027\\
40	0.000112915123278437\\
41	2.22264376967463e-05\\
42	0\\
43	0\\
44	0\\
45	0\\
46	0\\
47	0\\
48	0\\
49	0\\
50	0\\
};
\addlegendentry{End-to-end SER}

\end{axis}

\end{tikzpicture}%
\caption{Symbol error rate versus the number of training epochs for the closed-form, SVD, and end-to-end CV-NN SER optimization.}
\label{fig:ClosedVsCvx}
\end{figure}

Fig.~\ref{fig:ClosedVsCvx} compares our CV-NN solution against two baselines:
\begin{itemize}
\item {Closed-form} design that seeks per-stream alignment but is not jointly optimized \cite{gradientdecentSIMComm}, 
\item {CV-NN SVD-based} design that imposes orthogonality and attempts to diagonalize the channel {by using Alg. \eqref{alg:E2Etraining} with Frobenius cost function in \eqref{eq:GammaC_2}}. 
\end{itemize}

We plot the SER versus training epoch for a typical configuration with \(N_t = N_r = 484\) meta-atoms and \(L_t = L_r = 3\) layers. The closed-form method exhibits significant SER under dynamic channels, while the SVD approach stabilizes near \(0.09\) SER. In contrast, our method converges more rapidly and achieves near-zero SER, illustrating the advantage of non-orthogonal transformations discovered through gradient-based training. The main reason is that orthogonality cannot be maintained in time-varying environments.

\subsection{Performance Under Varying The Number of Elements and Layers}
\begin{figure}
\centering
\tikzset{every picture/.style={scale=.8}, every node/.style={scale=1}}
%
%

\definecolor{mycolor1}{rgb}{0.00000,0.44700,0.74100}%
\definecolor{mycolor2}{rgb}{0.85000,0.32500,0.09800}%
\begin{tikzpicture}

\begin{axis}[%
grid=major,
width=4.521in,
height=3.255in,
scale only axis,
xmin=49,
xmax=784,
xlabel style={at={(axis cs:416.5,0.0001)},anchor=north},
xlabel={Number of Elements $N$},
separate axis lines,
every outer y axis line/.append style={mycolor1},
ymin=0.0001,
ymax= 1,
ymode=log,
ylabel style={font=\color{black},at={(axis cs:49,0.01)},anchor=north},
ylabel={SER},
axis background/.style={fill=white},
xmajorgrids,
ymajorgrids,
legend style={{at =  (axis cs:49,0.0001)}, nodes={scale=0.75, transform shape}, anchor=south west, legend cell align=left, align=left, draw=black}
]

\addplot [color=blue, line width=1.5pt]
  table[row sep=crcr]{%
49	0.572500\\
64	0.4297\\
100	0.3564\\
169	0.2979\\
256	0.2448\\
361	0.1819\\
484	0.1238\\
625	0.0962\\
784	0.0905\\
};
\addlegendentry{SVD with $L_t = L_r = 2$}

\addplot [color=black, line width=1.5pt]
  table[row sep=crcr]{%
49	0.4028\\
64	0.3237\\
100	0.2232\\
169	0.1497\\
256	0.1110\\
361	0.0891\\
484	0.0768\\
625	0.0706\\
784	0.0668\\
};
\addlegendentry{SER with $L_t = L_r = 2$}

\addplot [color=blue, dashed, line width=1.5pt]
  table[row sep=crcr]{%
49	0.425000\\
64	0.3499\\
100	0.2885 \\
169	0.2166\\
256	0.1550\\
361	0.1136\\
484	0.0845\\
625	0.0611\\
784	0.0474\\
};
\addlegendentry{SVD with $L_t = L_r = 5$}

\addplot [color=black, dashed, line width=1.5pt]
  table[row sep=crcr]{%
49	0.3266\\
64	0.2298 \\
100	0.1414\\
169	0.0898\\
256	0.0597\\
361	0.0310\\
484	0.0098\\
625	0.0014\\
784	0.0001\\
};
\addlegendentry{SER with $L_t = L_r = 5$}

\end{axis}

\end{tikzpicture}%


\caption{Symbol error rate versus the number of elements for different number of SIS layers.}
\label{fig:M_sweep}
\end{figure}
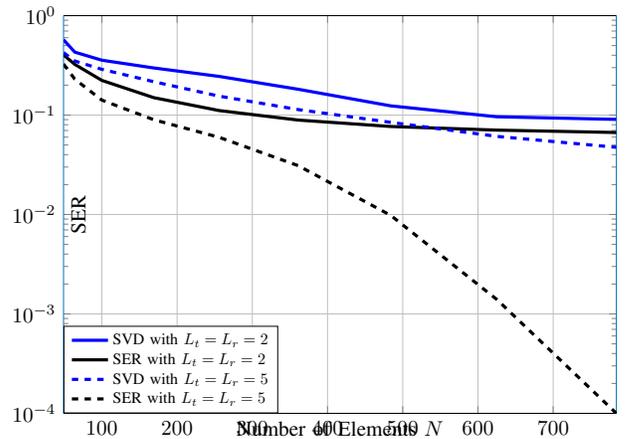

To assess scalability, we vary the SIS dimension \(N = N_t = N_r\) in Fig.~\ref{fig:M_sweep} and observe the final SER after convergence. While all schemes gain from a larger SIS, the gap between our CV-NN solution and the baselines widens. For instance, when the number of elements is 484 with 2 layers, the SVD-based method still lingers at a 0.09 SER, whereas the CV-NN optimization attains a SER of 0.05. For the same number of elements, we repeat the comparison with 5 SIS layers. Going from 2 to 5 layers, the SER drops with the number of elements to $10^{-4}$, thanks to the added wave-domain flexibility. Although the SVD solution also improves with more layers, it never matches the CV-NN learned solution. 

Overall, these results confirm that a {fully differentiable SIS} can achieve stable, low-SER solutions with modest training epochs, whereas conventional methods may fail to adapt or rely on assumptions (e.g., perfect orthogonality) that do not hold in practical 6G scenarios.

\section{Conclusion}
\label{conclusion}


In conclusion, we presented an end-to-end learning framework for stacked intelligent surfaces (SIS) by treating each element as a trainable complex neuron. Unlike conventional orthogonal or closed-form approaches, our gradient-based method {is trained for a representative channel distribution and can be fine-tuned whenever the channel statistics evolve, rather than re-optimizing for every instantaneous realization,} achieving lower error rates under diverse setups. Results confirm that allowing non-orthogonal wave transformations through deep complex-valued neural networks yields robust, scalable gains for multi-antenna systems. Future directions include exploring hardware constraints, multi-task objectives, and multi-user scenarios to further extend SIS capabilities in realistic deployments.

\vspace{-0.1cm}
\bibliographystyle{IEEEtran}
\bibliography{IEEEabrv,bibliography}

\end{document}